\documentclass[10pt]{article}
\usepackage{graphicx}
\usepackage{amsmath}
\usepackage{amssymb}
\usepackage{caption2}
\begin{document}
\bibliographystyle{prsty}
\begin{center}
{\large {\bf \sc{ Analysis of strong decays of the charmed mesons $D_J(2580)$, $D_J^*(2650)$, $D_J(2740)$,
$D^*_J(2760)$,  $D_J(3000)$, $D_J^*(3000)$ }}} \\[2mm]
Zhi-Gang Wang  \footnote{E-mail:zgwang@aliyun.com. } \\
  Department of Physics, North China Electric Power University, Baoding 071003, P. R.
  China
\end{center}

\begin{abstract}
In this article, we tentatively identify the  charmed mesons $D_J(2580)$, $D_J^*(2650)$,  $D_J(2740)$,
$D^*_J(2760)$,  $D_J(3000)$, $D_J^*(3000)$   observed by the LHCb collaboration according to their spin, parity and masses,
then study their strong decays  to the ground state charmed mesons plus light pseudoscalar mesons with
the heavy meson effective theory in the leading order approximation,  and obtain explicit expressions of the decay  widths.
The ratios among the decay widths can be used to confirm or reject   the   assignments of the   newly observed charmed mesons.
The strong coupling constants in the decay  widths can be fitted to the experimental
data in the future at the LHCb,  BESIII, KEK-B and $\rm{\bar{P}ANDA}$.
\end{abstract}


\section{Introduction}
Recently, the LHCb collaboration studied the $D^+\pi^-$, $D^0 \pi^+$, $D^{*+}\pi^-$ final states is $pp$ collisions at a center-of-mass energy
of $7\, \rm{TeV}$, observed the $D_1(2420)^0$  in the $D^{*+}\pi^-$ final state, the $D^*_2(2460)$ in the $D^+\pi^-$, $D^0 \pi^+$, $D^{*+}\pi^-$ final states, and
 measured  their parameters and confirmed their spin-parity assignments \cite{LHCb1307}.
    The LHCb collaboration also  observed two natural parity resonances $D_J^*(2650)^0$, $D_J^*(2760)^0$ and two unnatural parity resonances  $D_J(2580)^0$, $D_J(2740)^0$ in the $D^{*+}\pi^-$ mass spectrum,  and tentatively  identified the $D_J(2580)$ as the $2\, \rm{S}$ $0^-$ state,  the $D_J^*(2650)$ as the $2\, \rm{S}$ $1^-$ state, the  $D_J(2740)$  as the $1\, \rm {D}$ $2^-$ state, the $D_J^*(2760)$ as the $1\, \rm {D}$ $1^-$ state. The $D_J^*(2760)^0$ observed in the  $D^{*+} \pi^-$ and $D^+ \pi^-$ decay modes have  consistent parameters, their charged partner $D_J^*(2760)^+$ was observed in the $D^0 \pi^+$ final state \cite{LHCb1307}.
  Furthermore, the LHCb collaboration   observed one unnatural parity resonance $D_J(3000)^0$  in the $D^{*+} \pi^-$ final state,
two structures $D_J^*(3000)^0$ and $D_J^*(3000)^+$ in the $D^+ \pi^-$ and $D^0 \pi^+$ mass spectra, respectively \cite{LHCb1307}. The revelent parameters are presented in Table 1.

In 2010, the BaBar collaboration observed four excited  charmed mesons $D(2550)$, $D(2600)$, $D(2750)$ and $D(2760)$ in the decays
  $D^0(2550)\to D^{*+}\pi^-$, $D^{0}(2600)\to D^{*+}\pi^-,\,D^{+}\pi^-$,  $D^0(2750)\to D^{*+}\pi^-$,
$D^{0}(2760)\to D^{+}\pi^-$, $D^{+}(2600)\to D^{0}\pi^+$ and $D^{+}(2760)\to D^{0}\pi^+$ respectively in the inclusive $e^+e^-
\rightarrow c\bar{c}$  interactions  \cite{Babar2010}.  The BaBar collaboration also analyzed  the helicity distributions to determine
the spin-parity, and tentatively identified  the $(D(2550),D(2600))$  as  the $2\,\rm{S}$ doublet $(0^-,1^-)$,  the
$D(2750)$ and $D(2760)$ as the D-wave states. The revelent parameters are presented in Table 2, where we also present the possible correspondences among the particles observed by the LHCb and BaBar collaborations.

In Ref.\cite{Wang1009}, we study the strong decays of  the charmed mesons $D(2550)$, $D(2600)$, $D(2750)$ and $D(2760)$ with
the heavy meson effective theory in the leading order approximation, and tentatively identify the $(D(2550),D(2600))$ as the $2\,\rm{S}$ doublet
$(0^-,1^-)$, the $(D(2750),D(2760))$  as the $1\,\rm{D}$  doublet $(2^-,3^-)$, respectively.  Other studies lead to  similar or slightly different assignments \cite{ND-BaBar,Colangelo1207}. Now, we extend our previous work to study   the strong decays  of the  charmed mesons   observed by the LHCb collaboration with  the heavy meson  effective theory in the leading order approximation.

 \begin{table}
\begin{center}
\begin{tabular}{|cc|c|c|c|c|c|c }\hline\hline
                 &        & Mass [MeV]             & Width [MeV]              & Decay channel  & Significance \\ \hline
 $D_J^*(2650)^0$ &(N)     &$2649.2\pm3.5\pm3.5$    &$140.2\pm17.1\pm18.6$     &$D^{*+}\pi^-$   &24.5 $\sigma$       \\
 $D^*_J(2760)^0$ &(N)     &$2761.1\pm5.1\pm6.5$    &$74.4\pm3.4\pm37.0$       &$D^{*+}\pi^-$   &10.2 $\sigma$   \\
 $D_J(2580)^0$   &(U)     &$2579.5\pm3.4\pm5.5$    &$177.5\pm17.8\pm46.0$     &$D^{*+}\pi^-$   &18.8 $\sigma$ \\
 $D_J(2740)^0$   &(U)     &$2737.0\pm3.5\pm11.2$   &$73.2\pm13.4\pm 25.0$     &$D^{*+}\pi^-$   &7.2 $\sigma$   \\
 $D_J(3000)^0$   &(U)     &$2971.8\pm 8.7$         &$188.1\pm 44.8$           &$D^{*+}\pi^-$   &9.0 $\sigma$ \\ \hline
 $D_J^*(2760)^0$ &(N)     &$2760.1\pm1.1\pm3.7$    &$74.4\pm3.4\pm19.1$       &$D^+\pi^-$      &17.3 $\sigma$ \\
 $D_J^*(3000)^0$ &        &$3008.1 \pm 4.0$        &$110.5 \pm11.5$           &$D^+\pi^-$      &21.2 $\sigma$ \\ \hline
$D_J^*(2760)^+$  &        &$2771.7\pm 1.7\pm3.8$   &$66.7 \pm 6.6\pm10.5$     &$D^0\pi^+$      &18.8 $\sigma$\\
 $D_J^*(3000)^+$ &        &3008.1\, (fixed)        &110.5\, (fixed)           &$D^0\pi^+$      &6.6 $\sigma$ \\ \hline
 \hline
\end{tabular}
\end{center}
\caption{The experimental results from the LHCb collaboration, where the N and U denote the natural parity and unnatural parity, respectively.}
\end{table}

\begin{table}
\begin{center}
\begin{tabular}{|cc|c|c|c|c|c|c }\hline\hline
             &                               &Mass [MeV]                  &Width [MeV]          &Decay channel   \\
 $D^0(2550)$ &$\left[D_J(2580)^0\right]$     &$2539.4 \pm 4.5 \pm 6.8$    &$130\pm12 \pm13$     &$D^{*+}\pi^-$   \\
 $D^0(2600)$ &$\left[D^*_J(2650)^0\right]$   &$2608.7\pm 2.4\pm 2.5$      &$93\pm 6\pm13$       &$D^+\pi^-$,\,$D^{*+}\pi^-$ \\
 $D^0(2750)$ &$\left[D_J(2740)^0\right]$     &$2752.4\pm 1.7\pm 2.7$      &$71\pm6\pm11$        &$D^{*+}\pi^-$   \\
 $D^0(2760)$ &$\left[D^*_J(2760)^0\right]$   &$2763.3\pm 2.3\pm 2.3$      &$60.9\pm5.1\pm3.6$   &$D^{+}\pi^-$       \\
 $D^+(2600)$ &                               &$2621.3\pm 3.7\pm 4.2$      &$93$                 &$D^0\pi^+$ \\
 $D^+(2760)$ &$\left[D^*_J(2760)^+\right]$   &$2769.7\pm 3.8\pm 1.5$      &$60.9$               &$D^0\pi^+$    \\ \hline\hline
\end{tabular}
\end{center}
\caption{The experimental results from the BaBar collaboration, the particles in the bracket are the possible corresponding ones observed by the LHCb  collaboration.}
\end{table}

Let us take a short digression to discuss how to classify  the heavy-light  mesons. In the heavy quark limit, the heavy-light  mesons
$Q{\bar q}$  can be  classified in doublets according to the total
angular momentum of the light antiquark ${\vec s}_\ell$,
${\vec s}_\ell= {\vec s}_{\bar q}+{\vec L} $, where the ${\vec
s}_{\bar q}$ and ${\vec L}$ are the spin and orbital angular momentum of the light antiquark respectively \cite{RevWise}.
In the case of the radial quantum number $n=1$,
 the doublet $(P,P^*)$ have the spin-parity
$J^P_{s_\ell}=(0^-,1^-)_{\frac{1}{2}}$ for $L=0$;
 the two doublets $(P^*_0,P_1)$ and $(P_1,P^*_2)$ have the spin-parity
$J^P_{s_\ell}=(0^+,1^+)_{\frac{1}{2}}$ and $(1^+,2^+)_{\frac{3}{2}}$
respectively for $L=1$;  the  two doublets $(P^*_1,P_2)$ and
$(P_2,P_3^{ *})$ have the spin-parity
$J^P_{s_\ell}=(1^-,2^-)_{\frac{3}{2}}$ and $(2^-,3^-)_{\frac{5}{2}}$
respectively for $L=2$;  the  two doublets $(P^*_2,P_3)$ and
$(P_3,P_4^{ *})$ have the spin-parity
$J^P_{s_\ell}=(2^+,3^+)_{\frac{5}{2}}$ and $(3^+,4^+)_{\frac{7}{2}}$
respectively for $L=3$,
where the superscript $P$ denotes the parity. The
$n=2,3,4, \cdots$ states are clarified by analogous  doublets,
for example, $n=2$,  the  doublet $(P^{\prime},P^{*\prime})$
have the spin-parity $J^P_{s_\ell}=(0^-,1^-)_{\frac{1}{2}}$ for $L=0$.

 The helicity distributions from the  BaBar collaboration favor identifying the $D^0(2550)$ as the
$0^-$ state,  the $D^0(2600)$ as the $1^-$, $2^+$, $3^-$ state, and
the  $D^0(2750)$ as the $1^+$, $2^-$ state \cite{Babar2010}, which are compatible with the tentative assignments of the LHCb  collaboration \cite{LHCb1307}, see Table 2.
The $D_J(2580)^0$, $D_J(2740)^0$ and $D_J(3000)^0$ have unnatural parity,  and their possible spin-parity assignments  are $J^P=0^-,\,1^+,\,2^-,\,3^+,\,\cdots$.
The $D_J^*(2650)^0$ and $D_J^*(2760)^0$ have natural parity,  and their possible spin-parity assignments  are $J^P=0^+,\,1^-,\,2^+,\,3^-,\,\cdots$.
From the strong decays
\begin{eqnarray}
D_J^*(3000)^0&\to&D^{+}\pi^- \, , \nonumber\\
D_J^*(3000)^+&\to&D^{0}\pi^+ \, ,
\end{eqnarray}
we can conclude that the $D_J^*(3000)$ have the possible spin-parity $J^P=0^+,\,1^-,\,2^+,\,3^-,\,4^+,\cdots$.
The six low-lying states, $D$, $D^*$, $D_0(2400)$, $D_1(2430)$, $D_1(2420)$ and
$D_2(2460)$ are established \cite{PDG}, while the  $2\,\rm{S}$, $1\,\rm{D}$, $1\,\rm{F}$, $2\,\rm{P}$, and $3\,\rm{S}$ states are
still absent. The newly observed charmed mesons $D_J(2580)$, $D_J^*(2650)$, $D_J(2740)$, $D^*_J(2760)$, $D_J(3000)$, $D_J^*(3000)$
can be  tentatively  identified as
the missing $2\,\rm{S}$, $1\,\rm{D}$, $1\,\rm{F}$, $2\,\rm{P}$, and $3\,\rm{S}$ states.

The mass is a fundamental parameter in describing a hadron, in Table
3,  we present the predictions  from some theoretical models,  such
as the relativized quark model based on a universal
one-gluon exchange plus linear confinement potential
 \cite{GI},   the relativistic quark model includes the leading
order $1/M_h$ corrections \cite{PE}, the QCD-motivated relativistic
quark model based on the quasipotential approach \cite{EFG}. We can identify the $D_J(2580)$, $D_J^*(2650)$, $D_J(2740)$,
$D^*_J(2760)$, $D_J(3000)$, $D_J^*(3000)$ tentatively according to the masses.

In the following, we list out the possible assignments,
\begin{eqnarray}
(D_J(2580),D_J^*(2650))&=& (0^-,1^-)_{\frac{1}{2}}\, \, \,{\rm with} \,\,\, n=2, \,\, L=0\, ,\\
(D_J^*(2760),D_J(2740))&=& (1^-,2^-)_{\frac{3}{2}}\, \, \,{\rm with} \,\,\, n=1, \,\,L=2\, ,   \\
(D_J(2740),D_J^*(2760))&=& (2^-,3^-)_{\frac{5}{2}}\, \, \,{\rm with} \,\,\, n=1, \,\,L=2\, ,   \\
(D_J^*(3000),D_J(3000))&=& (2^+,3^+)_{\frac{5}{2}}  \, \, \,{\rm with} \,\,\,n=1, \,\,L=3\, , \\
(D_J(3000),D_J^*(3000))&=& (3^+,4^+)_{\frac{7}{2}}  \, \, \,{\rm with} \,\,\,n=1, \,\,L=3\, , \\
(D_J^*(3000),D_J(3000))&=& (0^+,1^+)_{\frac{1}{2}}  \, \, \,{\rm with} \,\,\,n=2, \,\,L=1\, , \\
(D_J(3000),D_J^*(3000))&=& (1^+,2^+)_{\frac{3}{2}}  \, \, \,{\rm with} \,\,\,n=2, \,\,L=1\, , \\
(D_J(3000),D_J^*(3000))&=& (0^-,1^-)_{\frac{1}{2}}  \, \, \,{\rm with} \,\,\,n=3, \,\,L=0\, .
\end{eqnarray}

In this work, we study the strong decays of the
charmed mesons $D_J(2580)$, $D_J^*(2650)$, $D_J(2740)$, $D^*_J(2760)$, $D_J(3000)$, $D_J^*(3000)$
observed by  the LHCb collaboration with  the heavy meson effective theory in the leading
order approximation, and make predictions for the decay widths and the ratios among the decay widths.
The ratios can be confronted with the experimental data in the future at the LHCb,  BESIII, KEK-B and $\rm{\bar{P}ANDA}$
 to distinguish the different assignments. Furthermore, the analytical expressions of the decay widths can be used to
determine the strong  coupling constants in the heavy mesons effective Lagrangian in the future. On the other hand, we can also use the ${}^3P_0$ model to
study those strong decays following Ref.\cite{ZhuShiLin}.
There have been several works using the heavy meson  effective theory
to identify the charmed  mesons \cite{Wang1009,Colangelo1207,Colangelo0511}, and
to study the radiative, vector-meson, two-pion decays
of the heavy quarkonium states \cite{HMET-RV}.

\begin{table}
\begin{center}
\begin{tabular}{|c|c|c|c|c|c|c|c|}\hline\hline
             & $n\,\,L\,\, s_\ell\,\,    J^P$        &Exp\cite{LHCb1307,PDG}   &GI\cite{LHCb1307,GI}   &PE\cite{PE} &EFG \cite{EFG}\\ \hline
   $D$       & $1\,\,{\rm S}\,\,\frac{1}{2}\,\,0^-$  &1867                     &1864                   &1868        &1871\\
   $D^*$     & $1\,\,{\rm S}\,\,\frac{1}{2}\,\,1^-$  &2008                     &2023                   &2005        &2010\\  \hline

   $D^*_0$   & $1\,\,{\rm P}\,\,\frac{1}{2}\,\,0^+$  &2400                     &2380                   &2377        &2406\\
   $D_1$     & $1\,\,{\rm P}\,\,\frac{1}{2}\,\,1^+$  &2427                     &2419                   &2490        &2469\\
   $D_1$     & $1\,\,{\rm P}\,\,\frac{3}{2}\,\,1^+$  &2420                     &2469                   &2417        &2426\\
   $D_2^*$   & $1\,\,{\rm P}\,\,\frac{3}{2}\,\,2^+$  &2460                     &2479                   &2460        &2460\\  \hline

   $D_1^*$   & $1\,\,{\rm D}\,\,\frac{3}{2}\,\,1^-$  &?\,2760\,(N)             &2796                   &2795        &2788  \\
   $D_2$     & $1\,\,{\rm D}\,\,\frac{3}{2}\,\,2^-$  &?\,2740\,(U)             &2801                   &2833        &2850  \\
   $D_2$     & $1\,\,{\rm D}\,\,\frac{5}{2}\,\,2^-$  &?\,2740\,(U)             &2806                   &2775        &2806 \\
   $D_3^*$   & $1\,\,{\rm D}\,\,\frac{5}{2}\,\,3^-$  &?\,2760\,(N)             &2806                   &2799        &2863  \\  \hline

   $D_2^*$   & $1\,\,{\rm F}\,\,\frac{5}{2}\,\,2^+$  &?\,3000\,(N)             &3074                   &3101        &3090  \\
   $D_3$     & $1\,\,{\rm F}\,\,\frac{5}{2}\,\,3^+$  &?\,3000\,(U)             &3074                   &3123        &3145  \\
   $D_3$     & $1\,\,{\rm F}\,\,\frac{7}{2}\,\,3^+$  &?\,3000\,(U)             &3079                   &3074        &3129 \\
   $D_4^*$   & $1\,\,{\rm F}\,\,\frac{7}{2}\,\,4^+$  &?\,3000\,(N)             &3084                   &3091        &3187  \\ \hline

   $D$       & $2\,\,{\rm S}\,\,\frac{1}{2}\,\,0^-$  &?\,2580\,(U)             &2558                   &2589        &2581\\
   $D^*$     & $2\,\,{\rm S}\,\,\frac{1}{2}\,\,1^-$  &?\,2650\,(N)             &2618                   &2692        &2632\\ \hline

   $D^*_0$   & $2\,\,{\rm P}\,\,\frac{1}{2}\,\,0^+$  & ?\,3000\,(N)            &                       &2949        &2919\\
   $D_1$     & $2\,\,{\rm P}\,\,\frac{1}{2}\,\,1^+$  & ?\,3000\,(U)            &                       &3045        &3021\\
   $D_1$     & $2\,\,{\rm P}\,\,\frac{3}{2}\,\,1^+$  & ?\,3000\,(U)            &                       &2995        &2932\\
   $D_2^*$   & $2\,\,{\rm P}\,\,\frac{3}{2}\,\,2^+$  & ?\,3000\,(N)            &                       &3035        &3012\\  \hline

   $D$       & $3\,\,{\rm S}\,\,\frac{1}{2}\,\,0^-$  &?\,3000\,(U)             &                       &3141        &3062\\
   $D^*$     & $3\,\,{\rm S}\,\,\frac{1}{2}\,\,1^-$  &?\,3000\,(N)             &                       &3226        &3096\\  \hline
  \hline
\end{tabular}
\end{center}
\caption{ The masses of the charmed mesons from different quark
models compared with experimental data, and the possible
assignments of the newly observed charmed mesons. The N and U denote the natural parity and unnatural parity, respectively.  }
\end{table}

The article is arranged as follows:  we study the strong decays of the
charmed mesons $D_J(2580)$, $D_J^*(2650)$, $D_J(2740)$, $D^*_J(2760)$, $D_J(3000)$, $D_J^*(3000)$
observed by  the LHCb collaboration   with the heavy meson 
effective theory in Sect.2; in Sect.3, we present the
 numerical results and discussions; and Sect.4 is reserved for our
conclusions.

\section{ The strong  decays with the heavy meson effective theory }

In the heavy meson effective theory,   the  spin doublets can be
described by the effective super-fields $H_a$, $S_a$,  $T_a$, $X_a$, $Y_a$, $Z_a$ and $R_a$,  respectively \cite{Falk1992},
\begin{eqnarray}
H_a & =& \frac{1+{\rlap{v}/}}{2}\left\{P_{a\mu}^*\gamma^\mu-P_a\gamma_5\right\} \, ,   \nonumber  \\
S_a &=& \frac{1+{\rlap{v}/}}{2} \left\{P_{1a}^{ \mu}\gamma_\mu\gamma_5-P_{0a}^*\right\}  \, , \nonumber \\
T_a^\mu &=&\frac{1+{\rlap{v}/}}{2} \left\{ P^{*\mu\nu}_{2a}\gamma_\nu-P_{1a\nu} \sqrt{3 \over 2} \gamma_5 \left[ g^{\mu \nu}-{\gamma^\nu
(\gamma^\mu-v^\mu) \over 3} \right]\right\}\, ,   \nonumber\\
X_a^\mu &=&\frac{1+{\rlap{v}/}}{2} \Bigg\{ P^{\mu\nu}_{2a}
\gamma_5\gamma_\nu -P^{*}_{1a\nu} \sqrt{3 \over 2} \left[ g^{\mu \nu}-{\gamma^\nu (\gamma^\mu+v^\mu) \over 3}  \right]\Bigg\} \, , \nonumber  \\
Y_a^{ \mu\nu} &=&\frac{1+{\rlap{v}/}}{2} \left\{P^{*\mu\nu\sigma}_{3a} \gamma_\sigma -P^{\alpha\beta}_{2a}\sqrt{5 \over 3} \gamma_5 \left[ g^\mu_\alpha g^\nu_\beta -{g^\nu_\beta\gamma_\alpha  (\gamma^\mu-v^\mu) \over 5} - {g^\mu_\alpha\gamma_\beta  (\gamma^\nu-v^\nu) \over 5}  \right]\right\}\, ,\nonumber\\
Z_a^{ \mu\nu} &=&\frac{1+{\rlap{v}/}}{2} \left\{P^{\mu\nu\sigma}_{3a}\gamma_5 \gamma_\sigma -P^{*\alpha\beta}_{2a}\sqrt{5 \over 3}  \left[ g^\mu_\alpha g^\nu_\beta -{g^\nu_\beta\gamma_\alpha  (\gamma^\mu+v^\mu) \over 5} - {g^\mu_\alpha\gamma_\beta  (\gamma^\nu+v^\nu) \over 5}  \right]\right\}\, ,\nonumber\\
R_a^{ \mu\nu\rho} &=&\frac{1+{\rlap{v}/}}{2} \left\{P^{*\mu\nu\rho\sigma}_{4a}\gamma_5 \gamma_\sigma -P^{\alpha\beta\tau}_{3a}\sqrt{7 \over 4}  \left[ g^\mu_\alpha g^\nu_\beta g^\rho_\tau-{g^\nu_\beta g^\rho_\tau\gamma_\alpha (\gamma^\mu-v^\mu) \over 7} - {g^\mu_\alpha g^\rho_\tau\gamma_\beta  (\gamma^\nu-v^\nu) \over 7}
\right.\right.\nonumber\\
&&\left.\left.- {g^\mu_\alpha g^\nu_\beta\gamma_\tau  (\gamma^\rho-v^\rho) \over 7} \right]\right\}\, ,
\end{eqnarray}
where the  heavy meson fields  $P^{(*)}$ contain a factor $\sqrt{M_{P^{(*)}}}$ and
have dimension of mass $\frac{3}{2}$.
The super-fields $H_a$ contain the $\rm{S}$-wave mesons, $S_a$, $T_a$ contain  the $\rm{P}$-wave mesons, $X_a$, $Y_a$ contain  the $\rm{D}$-wave mesons, $Z_a$ and $R_a$ contain  the $\rm{F}$-wave mesons.
The $n=1,\,2,\,3, \cdots$  heavy mesons with the same heavy flavor have the same
 parity, time-reversal and charge conjunction properties except
for the masses, and can be combined into the super-fields: $H_a$,
$H_a'$, $H_a''$, $\cdots$; $S_a$, $S_a'$, $S_a''$, $\cdots$; $T_a$,
$T_a'$, $T_a''$, $\cdots$; etc, where the superscripts  $\prime$,
$\prime\prime$ and $\prime\prime\prime$ denote the $n=2,\,3, \, 4, \cdots$  states, respectively. We can
replace  the heavy meson fields $P^{(*)}$ with  their corresponding radial excited states to  obtain
the corresponding super-fields $H_a^\prime$, $S_a^\prime$, $\cdots$.

 The light pseudoscalar mesons are described by the fields
 $\displaystyle \xi=e^{i {\cal M} \over
f_\pi}$, where
\begin{equation}
{\cal M}= \left(\begin{array}{ccc}
\sqrt{\frac{1}{2}}\pi^0+\sqrt{\frac{1}{6}}\eta & \pi^+ & K^+\nonumber\\
\pi^- & -\sqrt{\frac{1}{2}}\pi^0+\sqrt{\frac{1}{6}}\eta & K^0\\
K^- & {\bar K}^0 &-\sqrt{\frac{2}{3}}\eta
\end{array}\right) \, ,
\end{equation}
and $f_\pi=130\,\rm{MeV}$.

At the leading order, the heavy meson chiral Lagrangians ${\cal L}_0$, ${\cal
L}_H$, ${\cal L}_S$, ${\cal L}_T$, ${\cal L}_X$, ${\cal L}_Y$, ${\cal L}_Z$, ${\cal L}_R$  for
the strong decays to the $D^{(*)}\pi$, $D^{(*)}\eta$ and $D_s^{(*)}K$ states can be  written as:
\begin{eqnarray}
{\cal L}_0&=&i{\rm Tr} \left\{{\bar H}_a {v\cdot\cal D}_{ab} H_b     \right\} +i{\rm Tr} \left\{{\bar S}_a {v\cdot\cal D}_{ab} S_b\right\}
+i{\rm Tr} \left\{{\bar T}^\mu_a {v\cdot\cal D}_{ab} T_{\mu b}   \right\} +i{\rm Tr} \left\{{\bar X}^\mu_a {v\cdot\cal D}_{ab} X_{\mu b}   \right\}
\nonumber \\
&&+i{\rm Tr} \left\{{\bar Y}^{\mu\nu}_a {v\cdot\cal D}_{ab} Y_{\mu\nu b}   \right\}
 +i{\rm Tr} \left\{{\bar Z}^{\mu\nu}_a {v\cdot\cal D}_{ab} Z_{\mu\nu b}   \right\}+i{\rm Tr} \left\{{\bar R}^{\mu\nu\alpha}_a {v\cdot\cal D}_{ab} Z_{\mu\nu\alpha b}  \right\}\nonumber \\
 &&-\delta m_S{\rm Tr} \left\{{\bar S}_a  S_a\right\}-\delta m_T{\rm Tr} \left\{{\bar T}^\mu_a  T_{\mu a}   \right\} -\delta m_X {\rm Tr} \left\{{\bar X}^\mu_a X_{\mu a}   \right\}-\delta m_Y{\rm Tr} \left\{{\bar Y}^{\mu\nu}_a  Y_{\mu\nu a}   \right\} \nonumber\\
&&-\delta m_Z{\rm Tr} \left\{{\bar Z}^{\mu\nu}_a  Z_{\mu\nu a}   \right\}-\delta m_R{\rm Tr} \left\{{\bar R}^{\mu\nu\beta}_a  Z_{\mu\nu\beta a}   \right\}\, ,
\nonumber \\
{\cal L}_H &=& \,  g_H {\rm Tr} \left\{{\bar H}_a H_b \gamma_\mu\gamma_5 {\cal A}_{ba}^\mu \right\} \, ,\nonumber \\
{\cal L}_S &=& \,  g_S {\rm Tr} \left\{{\bar H}_a S_b \gamma_\mu \gamma_5 {\cal A}_{ba}^\mu \right\}\, + \, h.c. \, , \nonumber \\
{\cal L}_T &=&  {g_T \over \Lambda}{\rm Tr}\left\{{\bar H}_aT^\mu_b (i {\cal D}_\mu {\not\! {\cal A}  }+i{\not\! {\cal D}  } { \cal A}_\mu)_{ba} \gamma_5\right\} + h.c. \, , \nonumber \\
{\cal L}_X &=& {g_X \over \Lambda}{\rm Tr}\left\{{\bar H}_a X^\mu_b(i {\cal D}_\mu {\not\! {\cal A}  }+i{\not\! {\cal D}  } { \cal A}_\mu)_{ba} \gamma_5\right\} + h.c. \, ,\nonumber   \\
{\cal L}_{Y} &=&  {1 \over {\Lambda^2}}{\rm Tr}\left\{ {\bar H}_a Y^{\mu \nu}_b \left[k_1^Y \{{\cal D}_\mu, {\cal D}_\nu\} {\cal A}_\lambda + k_2^Y \left({\cal D}_\mu {\cal D}_\lambda { \cal A}_\nu + {\cal D}_\nu {\cal D}_\lambda { \cal A}_\mu \right)\right]_{ba}  \gamma^\lambda \gamma_5\right\} + h.c. \, , \nonumber\\
{\cal L}_{Z} &=&  {1 \over {\Lambda^2}}{\rm Tr}\left\{ {\bar H}_a Z^{\mu \nu}_b \left[k_1^Z \{ {\cal D}_\mu, {\cal D}_\nu\} {\cal A}_\lambda + k_2^Z \left({\cal D}_\mu {\cal D}_\lambda { \cal A}_\nu + {\cal D}_\nu {\cal D}_\lambda { \cal A}_\mu \right)\right]_{ba}  \gamma^\lambda \gamma_5\right\} + h.c. \, , \nonumber\\
{\cal L}_{R} &=&{1 \over {\Lambda^3}}{\rm Tr}\left\{ {\bar H}_a R^{\mu \nu \rho}_b \left[k_1^R \{{\cal D}_\mu, {\cal D}_\nu, {\cal D}_\rho\} {\cal A}_\lambda + k_2^R \left(\{{\cal D}_\mu,{\cal D}_\rho\} {\cal D}_\lambda { \cal A}_\nu + \{{\cal D}_\nu,{\cal D}_\rho\} {\cal D}_\lambda { \cal A}_\mu     \right.\right.\right.\nonumber\\
&&\left.\left.\left.+ \{{\cal D}_\mu,{\cal D}_\nu\} {\cal D}_\lambda { \cal A}_\rho\right)\right]_{ba}  \gamma^\lambda \gamma_5\right\} + h.c. \, ,
  \end{eqnarray}
where
\begin{eqnarray}
{\cal D}_{\mu}&=&\partial_\mu+{\cal V}_{\mu} \, , \nonumber \\
 {\cal V}_{\mu }&=&\frac{1}{2}\left(\xi^\dagger\partial_\mu \xi+\xi\partial_\mu \xi^\dagger\right)\, , \nonumber \\
 {\cal A}_{\mu }&=&\frac{1}{2}\left(\xi^\dagger\partial_\mu \xi-\xi\partial_\mu  \xi^\dagger\right)\,  , \nonumber\\
 \{ {\cal D}_\mu, {\cal D}_\nu \}&=&{\cal D}_\mu {\cal D}_\nu+{\cal D}_\nu {\cal D}_\mu  \, ,\nonumber\\
 \{ {\cal D}_\mu, {\cal D}_\nu, {\cal D}_\rho \}&=&{\cal D}_\mu {\cal D}_\nu{\cal D}_\rho+{\cal D}_\mu {\cal D}_\rho{\cal D}_\nu+{\cal D}_\nu {\cal D}_\mu{\cal D}_\rho +{\cal D}_\nu {\cal D}_\rho{\cal D}_\mu+{\cal D}_\rho{\cal D}_\mu {\cal D}_\nu+{\cal D}_\rho{\cal D}_\nu {\cal D}_\mu \, ,
\end{eqnarray}
  $\delta m_S=m_S-m_H$, $\delta m_T=m_T-m_H$,  $\delta m_X=m_X-m_H$, $\delta m_Y=m_Y-m_H$, $\delta m_Z=m_Z-m_H$, $\delta m_R=m_R-m_H$, $\Lambda$ is the chiral symmetry-breaking scale and taken as
$\Lambda = 1 \, \rm{GeV} $  \cite{Colangelo0511}, the strong coupling
constants  $g_H$, $g_S$, $g_T$, $g_X$, $g_Y=k^Y_1+k^Y_2$, $g_Z=k^Z_1+k^Z_2$ and $g_R=k^R_1+k^R_2$ can be
fitted  to the experimental data.
The heavy meson chiral Lagrangians  ${\cal
L}_H$, ${\cal L}_S$, ${\cal L}_T$, ${\cal L}_X$ and ${\cal L}_Y$  are taken from Ref.\cite{HL-1}, the  ${\cal L}_Z$ and ${\cal L}_R$
are constructed accordingly  in this article.
The subscript indexes $H$, $S$, $T$, $X$, $Y$, $Z$ and $R$ denote the
interactions between the super-field $H$ and the super-fields $H$,
$S$, $T$, $X$, $Y$, $Z$ and $R$, respectively. We  smear the
superscripts $\prime$, $\prime\prime$, $\prime\prime\prime$,
$\cdots$ for simplicity, the notation $g_H$ denotes the strong
coupling constants in the vertexes $HH{\cal A}$, $H'H{\cal A}$,
$H'H'{\cal A}$, $H''H{\cal A}$, $\cdots$, the notations $g_S$,
$g_T$, $g_X$, $g_Y$, $g_Z$ and $g_R$ should be  understood in the same way.
We can also study the decays to the light vector mesons $V$ besides the pseudoscalar mesons $\mathcal{P}$ with the  replacement ${\cal V}_{\mu} \to {\cal V}_{\mu}+V_\mu$,  and introduce additional phenomenological Lagrangians \cite{PRT1997}, therefore additional unknown
coupling constants, which have to be fitted to the precise experimental data in the future, and is beyond the present work.

From the heavy meson chiral Lagrangians  ${\cal L}_H$, ${\cal L}_S$,
${\cal L}_T$, ${\cal L}_X$, ${\cal L}_Y$, ${\cal L}_Z$ and ${\cal L}_R$, we can obtain the  widths
$\Gamma$ for strong decays to the final states $D^{(*)}\pi$, $D^{(*)}\eta$ and
$D_s^{(*)}K$,
\begin{eqnarray}
\Gamma&=&\frac{1}{2J+1}\sum\frac{p_{f}}{8\pi M^2_i } |T|^2\, , \nonumber\\
p_f&=&\frac{\sqrt{(M_i^2-(M_f+m_\mathcal{P})^2)(M_i^2-(M_f-m_\mathcal{P})^2)}}{2M_i}\, ,
\end{eqnarray}
where the $T$ denotes the scattering amplitudes,  the $i$ and $f$ denote the initial and final state heavy mesons, respectively, the  $J$ is the total angular momentum  of the initial heavy meson, the $\sum$ denotes the summation of all the  polarization vectors of the total angular momentum $j=1$, $2$, $3$ or $4$, and the $\mathcal{P}$ denotes the light pseudoscalar mesons.

Now we write down the explicit expressions of the decay widths $\Gamma$ in different channels,

$\bullet$  $(0^-,1^-)_{\frac{1}{2}}\to (0^-,1^-)_{\frac{1}{2}}+ \mathcal{P}$,
\begin{eqnarray}
\Gamma(1^- \to 1^-) &=&C_{\mathcal{P}} \frac{g_H^2M_f p_f^3}{3\pi f_{\pi}^2M_i} \, \left[C_{\mathcal{P}} \frac{g_H^2(M_i+M_f)^2p_f^3}{12\pi f_{\pi}^2M_i M_f} \right]\, , \\
\Gamma(1^- \to 0^-) &=&C_{\mathcal{P}} \frac{g_H^2M_f p_f^3}{6\pi f_{\pi}^2M_i} \, \left[C_{\mathcal{P}} \frac{g_H^2(M_i+M_f)^2p_f^3}{24\pi f_{\pi}^2M_i M_f} \right]\, , \\
\Gamma(0^- \to 1^-) &=&C_{\mathcal{P}} \frac{g_H^2M_f p_f^3}{2\pi f_{\pi}^2M_i} \, \left[C_{\mathcal{P}} \frac{g_H^2(M_i+M_f)^2p_f^3}{8\pi f_{\pi}^2M_i M_f} \right] \, ,
\end{eqnarray}

$\bullet$  $(0^+,1^+)_{\frac{1}{2}}\to (0^-,1^-)_{\frac{1}{2}}+\mathcal{P}$,
\begin{eqnarray}
\Gamma(1^+ \to 1^-) &=&C_{\mathcal{P}} \frac{g_S^2M_f \left(p_f^2 + m_\mathcal{P}^2 \right)p_f}{2\pi f_{\pi}^2 M_i } \,  \left[C_{\mathcal{P}} \frac{g_S^2 (M_i-M_f)^2\left(p_f^2 + m_\mathcal{P}^2 \right)p_f}{8\pi f_{\pi}^2 M_i M_f } \right]\, , \\
\Gamma(0^+ \to 0^-) &=&C_{\mathcal{P}} \frac{g_S^2M_f \left(p_f^2 + m_\mathcal{P}^2 \right)p_f}{2\pi f_{\pi}^2 M_i } \,  \left[C_{\mathcal{P}} \frac{g_S^2 (M_i-M_f)^2\left(p_f^2 + m_\mathcal{P}^2 \right)p_f}{8\pi f_{\pi}^2 M_i M_f } \right]\, ,
\end{eqnarray}

$\bullet$  $(1^+,2^+)_{\frac{3}{2}}\to (0^-,1^-)_{\frac{1}{2}}+ \mathcal{P}$,
\begin{eqnarray}
\Gamma(2^+ \to 1^-) &=&C_{\mathcal{P}} \frac{2g_T^2M_f p_f^5}{5\pi f_{\pi}^2\Lambda^2 M_i} \, \left[C_{\mathcal{P}} \frac{g_T^2(M_i+M_f)^2 p_f^5}{10\pi f_{\pi}^2\Lambda^2 M_i M_f} \right]\, , \\
\Gamma(2^+ \to 0^-) &=&C_{\mathcal{P}} \frac{4g_T^2M_f p_f^5}{15\pi f_{\pi}^2\Lambda^2 M_i} \, \left[C_{\mathcal{P}} \frac{g_T^2(M_i+M_f)^2 p_f^5}{15\pi f_{\pi}^2\Lambda^2 M_i M_f} \right]\, ,\\
\Gamma(1^+ \to 1^-) &=&C_{\mathcal{P}} \frac{2g_T^2M_f p_f^5}{3\pi f_{\pi}^2 \Lambda^2 M_i } \, \left[C_{\mathcal{P}} \frac{g_T^2(M_i+M_f)^2 p_f^5}{6\pi f_{\pi}^2\Lambda^2 M_i M_f} \right]\, ,
\end{eqnarray}

$\bullet$  $(1^-,2^-)_{\frac{3}{2}}\to (0^-,1^-)_{\frac{1}{2}}+ \mathcal{P}$,
\begin{eqnarray}
\Gamma(2^- \to 1^-) &=&C_{\mathcal{P}} \frac{2g_X^2M_f\left(p_f^2+m_\mathcal{P}^2\right) p_f^3}{3\pi f_{\pi}^2\Lambda^2 M_i } \, \left[C_{\mathcal{P}} \frac{g_X^2(M_i-M_f)^2\left(p_f^2+m_\mathcal{P}^2\right) p_f^3}{6\pi f_{\pi}^2\Lambda^2 M_i M_f} \right]\, , \\
\Gamma(1^- \to 1^-) &=&C_{\mathcal{P}} \frac{2g_X^2M_f\left(p_f^2+m_\mathcal{P}^2\right) p_f^3}{9\pi f_{\pi}^2\Lambda^2 M_i } \, \left[C_{\mathcal{P}} \frac{g_X^2(M_i-M_f)^2\left(p_f^2+m_\mathcal{P}^2\right) p_f^3}{18\pi f_{\pi}^2\Lambda^2 M_i M_f} \right]\, , \\
\Gamma(1^- \to 0^-) &=&C_{\mathcal{P}} \frac{4g_X^2M_f\left(p_f^2+m_\mathcal{P}^2\right) p_f^3}{9\pi f_{\pi}^2\Lambda^2 M_i } \, \left[C_{\mathcal{P}} \frac{g_X^2(M_i-M_f)^2\left(p_f^2+m_\mathcal{P}^2\right) p_f^3}{9\pi f_{\pi}^2\Lambda^2 M_i M_f} \right]\, ,
\end{eqnarray}

$\bullet$  $(2^-,3^-)_{\frac{5}{2}}\to (0^-,1^-)_{\frac{1}{2}}+ \mathcal{P}$,
\begin{eqnarray}
\Gamma(3^- \to 1^-) &=&C_{\mathcal{P}} \frac{16g_Y^2M_f p_f^7}{105\pi f_{\pi}^2\Lambda^4 M_i}\, \left[C_{\mathcal{P}} \frac{4g_Y^2(M_i+M_f)^2 p_f^7}{105\pi f_{\pi}^2\Lambda^4 M_i M_f} \right] \, , \\
\Gamma(3^- \to 0^-) &=&C_{\mathcal{P}} \frac{4g_Y^2M_f p_f^7}{35\pi f_{\pi}^2\Lambda^4 M_i} \, \left[C_{\mathcal{P}} \frac{g_Y^2(M_i+M_f)^2 p_f^7}{35\pi f_{\pi}^2\Lambda^4 M_i M_f} \right]\, , \\
\Gamma(2^- \to 1^-) &=&C_{\mathcal{P}} \frac{4g_Y^2M_f p_f^7}{15\pi f_{\pi}^2\Lambda^4 M_i} \, \left[C_{\mathcal{P}} \frac{g_Y^2(M_i+M_f)^2 p_f^7}{15\pi f_{\pi}^2\Lambda^4 M_i M_f} \right]\, ,
\end{eqnarray}

$\bullet$  $(2^+,3^+)_{\frac{5}{2}}\to (0^-,1^-)_{\frac{1}{2}}+ \mathcal{P}$,
\begin{eqnarray}
\Gamma(3^+ \to 1^-) &=&C_{\mathcal{P}} \frac{4g_Z^2M_f\left( p_f^2+m_\mathcal{P}^2\right) p_f^5}{15\pi f_{\pi}^2\Lambda^4 M_i }\,\left[ C_{\mathcal{P}} \frac{g_Z^2(M_i-M_f)^2\left( p_f^2+m_\mathcal{P}^2\right) p_f^5}{15\pi f_{\pi}^2\Lambda^4 M_i M_f}\right] \, , \\
\Gamma(2^+ \to 1^-) &=&C_{\mathcal{P}} \frac{8g_Z^2M_f\left( p_f^2+m_\mathcal{P}^2\right) p_f^5}{75\pi f_{\pi}^2\Lambda^4 M_i }\,\left[ C_{\mathcal{P}} \frac{2g_Z^2(M_i-M_f)^2\left( p_f^2+m_\mathcal{P}^2\right) p_f^5}{75\pi f_{\pi}^2\Lambda^4 M_i M_f}\right] \, ,\\
\Gamma(2^+ \to 0^-) &=&C_{\mathcal{P}} \frac{4g_Z^2M_f \left(p_f^2+m_\mathcal{P}^2 \right) p_f^5}{25\pi f_{\pi}^2 \Lambda^4 M_i  } \,\left[ C_{\mathcal{P}} \frac{g_Z^2(M_i-M_f)^2\left( p_f^2+m_\mathcal{P}^2\right) p_f^5}{25\pi f_{\pi}^2\Lambda^4 M_i M_f}\right] \, ,
\end{eqnarray}

$\bullet$  $(3^+,4^+)_{\frac{7}{2}}\to (0^-,1^-)_{\frac{1}{2}}+ \mathcal{P}$,
\begin{eqnarray}
\Gamma(4^+ \to 1^-) &=&C_{\mathcal{P}} \frac{4g_R^2M_f p_f^9}{7\pi f_{\pi}^2\Lambda^6 M_i} \, \left[C_{\mathcal{P}} \frac{g_R^2(M_i+M_f)^2 p_f^9}{7\pi f_{\pi}^2\Lambda^6 M_i M_f} \right]\, , \\
\Gamma(4^+ \to 0^-) &=&C_{\mathcal{P}} \frac{16g_R^2M_f p_f^9}{35\pi f_{\pi}^2\Lambda^6 M_i} \, \left[C_{\mathcal{P}} \frac{4g_R^2(M_i+M_f)^2 p_f^9}{35\pi f_{\pi}^2\Lambda^6 M_i M_f} \right]\, , \\
\Gamma(3^+ \to 1^-) &=&C_{\mathcal{P}} \frac{36g_R^2M_f p_f^9}{35\pi f_{\pi}^2\Lambda^6 M_i} \, \left[C_{\mathcal{P}} \frac{9g_R^2(M_i+M_f)^2 p_f^9}{35\pi f_{\pi}^2\Lambda^6 M_i M_f} \right] \, ,
\end{eqnarray}
 the coefficients $C_{\pi^\pm}=C_{K^\pm}=C_{K^0}=C_{\bar{K}^0}=1$, $C_{\pi^0}=\frac{1}{2}$ and $C_{\eta}=\frac{2}{3}$. We obtain the expressions in
  the bracket $"\left[ \right]"$ by taking into account the different four-velocities of the initial and final state heavy mesons. The on-shell conditions require
  that $M_iv_\mu=M_fv_\mu^{\prime}+p_\mu$ with $v^2=v^{\prime2}=1$, $p^2=m_\mathcal{P}^2$, $v_\mu\neq v^{\prime}_\mu$. In some multiplets, the conditions $v_\mu= v^{\prime}_\mu$ and $v_\mu\neq v^{\prime}_\mu$ lead to quite different decay widths but similar ratios among the decay widths. The ratios among the decay widths can be used to identify the heavy mesons, we expect that the expressions in the bracket $"\left[ \right]"$ cannot lead to different conclusions.
In calculations, we take  the approximation ${\cal{A}}_\mu\approx i\frac{\partial_\mu {\cal{M}}}{f_{\pi}} $ and neglect
the intermediate loops of   light pseudoscalar mesons.
Furthermore, we neglect the flavor and spin violation corrections of order
$\mathcal {O}(1/m_Q)$ to avoid introducing  new unknown coupling constants,
 and we expect that the corrections would not be larger
than (or as large as) the leading order contributions.

\section{Numerical Results}
The input parameters are taken  as
$M_{\pi^+}=139.57\,\rm{MeV}$, $M_{\pi^0}=134.9766\,\rm{MeV}$,
$M_{K^+}=493.677\,\rm{MeV}$, $M_{\eta}=547.853\,\rm{MeV}$,
$M_{D^+}=1869.60\,\rm{MeV}$, $M_{D^0}=1864.83\,\rm{MeV}$,
$M_{D_s^+}=1968.47\,\rm{MeV}$, $M_{D^{*+}}=2010.25\,\rm{MeV}$,
$M_{D^{*0}}=2006.96\,\rm{MeV}$, $M_{D_s^{*+}}=2112.3\,\rm{MeV}$ \cite{PDG},
 $M_{D(2420)}=2419.6\,\rm{MeV}$, $M_{D_2^*(2460)}=2460.4\,\rm{MeV}$ \cite{LHCb1307}.

 The numerical values of the widths  of the strong
 decays of the charmed mesons $D_J(2580)$, $D_J^*(2650)$, $D_J(2740)$, $D^*_J(2760)$,
 $D_J(3000)$, $D_J^*(3000)$ observed by the LHCb  collaboration
  are presented in Tables 4-5, where we retain the strong coupling constants $g_H$, $g_X$, $g_Y$, $g_Z$ and $g_R$.
The strong coupling constants can be fitted to the experimental
data in the future at the LHCb,  BESIII, KEK-B and $\rm{\bar{P}ANDA}$, and  taken as basic input parameters in studying the interactions among the heavy mesons.
The strong coupling constant $g_H$ for $n=1$ vary  in
a large range $g_H=0.1-0.6$ from different theoretical approaches, it is difficult to choose the optimal  value \cite{Wang2007}, we usually
fit the $g_H$ to  the decay width  $\Gamma(D^{*+} \to D^0 \pi^+)$ from the CLEO collaboration
\cite{CLEO-gH1,CLEO-gH2}. The strong coupling
constants $g_H$ (with $n=2,3$), $g_S$ (with $n=2$), $g_T$ (with $n=2$), $g_X$, $g_Y$, $g_Z$, $g_R$ involve the radial excited S-wave and P-wave heavy mesons and
ground state D-wave and F-wave heavy mesons,  it is impossible to determine their values
with the heavy quark (or meson) effective theory itself without enough
experimental data. The existing theoretical works focus on the strong
coupling constants $g_H$, $g_S$, $g_T$ for the ground state S-wave
and P-wave heavy mesons \cite{PRT1997,Wang2007,Wang2006}, while
the works  on other strong coupling constants are rare  \cite{Zhu1003}.

In Table 6-7,  we present the ratios $\widehat{\Gamma}=\frac{\Gamma}{\Gamma(D^{(*)}_J\to D^{*+}\pi^-)}$ of the strong decays of the   charmed
mesons $D_J(2580)$, $D_J^*(2650)$, $D_J(2740)$, $D^*_J(2760)$, $D_J(3000)$, $D_J^*(3000)$
observed by  the LHCb collaboration, which can be used to identify the charmed mesons by confronting them with the experimental data in the future.
In previous work \cite{Wang1009},  we tentatively identify the
$(D(2550),D(2600))$ as the doublet $(0^-,1^-)_{\frac{1}{2}}$ with
$n=2$,  the $(D(2750),D(2760))$ as the doublet $(2^-,3^-)_{\frac{5}{2}}$ with
$n=1$ via analyzing the  ratios of the branching fractions,
\begin{eqnarray}
\frac{{\rm{Br}}\left(D^*_2(2460)^0\to D^+\pi^-\right)}{{\rm{Br}}\left(D^*_2(2460)^0\to D^{*+}\pi^-\right)}  \, , \nonumber \\
\frac{{\rm{Br}}\left(D(2600)^0\to D^+\pi^-\right)}{{\rm{Br}}\left(D(2600)^0\to D^{*+}\pi^-\right)} \, , \nonumber \\
\frac{{\rm{Br}}\left(D(2760)^0\to D^+\pi^-\right)}{{\rm{Br}}\left(D(2750)^0\to D^{*+}\pi^-\right)} \, ,
\end{eqnarray}
with the heavy meson  effective theory in the leading order approximation. The measurement of the LHCb collaboration also
favors the assignment $(D_J(2580),D_J^*(2650))= (0^-,1^-)_{\frac{1}{2}}$  with $n=2$ \cite{LHCb1307}.
The helicity distribution from the BaBar collaboration
 disfavors identifying the $D(2750)$ as the $3^-$ state
\cite{Babar2010}, which is compatible with the measurement of the LHCb collaboration that the $D_J(2740)$ has unnatural parity \cite{LHCb1307}.
The measurement of the LHCb collaboration favors two possible assignments,
\begin{eqnarray}
(D_J^*(2760),D_J(2740))&=& (1^-,2^-)_{\frac{3}{2}}\, \, \,{\rm with} \,\,\, n=1, \,\,L=2\, ,   \\
(D_J(2740),D_J^*(2760))&=& (2^-,3^-)_{\frac{5}{2}}\, \, \,{\rm with} \,\,\, n=1, \,\,L=2\,  .
\end{eqnarray}
 We tentatively identify
the $D_J(2740)$ as the $1\,\rm{D}$ state with $J^P=2^-$, however, the assignments  $1\,D\,\frac{3}{2}\,2^-$ and  $1\,D\,\frac{5}{2}\,2^-$ lead to quite different
ratios among the decay widths. The $D_J^*(2760)$ have natural parity, the assignments  $1\,D\,\frac{3}{2}\,1^-$ and  $1\,D\,\frac{5}{2}\,3^-$ also  lead to quite different ratios among the decay widths. We can confront the present predictions  with the experimental data in the future to identify the  newly observed charmed mesons.

\begin{table}
\begin{center}
\begin{tabular}{|c|c|cc|cc| }\hline\hline
                 & $n\,L\,s_\ell\,J^P$         & Decay channels    & Widths  [GeV]      &Decay channels     & Widths [GeV]   \\ \hline

 $D_J(2580)$     & $2\,S\,\frac{1}{2}\,0^-$    & $D^{*+}\pi^-$     & $0.86744\,g_H^2$    &                  &    \\
                 &                             & $D^{*+}_sK^-$     & 0                   &                  &   \\
                 &                             & $D^{*0}\pi^0$     & $0.44284\,g_H^2$    &                  &   \\
                 &                             & $D^{*0}\eta$      & $0.01545\,g_H^2$    &                  &   \\  \hline

 $D^*_J(2650)$   & $2\,S\,\frac{1}{2}\,1^-$    & $D^{*+}\pi^-$     & $0.78436\,g_H^2$    & $D^{+}\pi^-$     & $0.61923\,g_H^2$   \\
                 &                             & $D^{*+}_sK^-$     & $0.03361\,g_H^2$    & $D_s^{+} K^-$    & $0.15626\,g_H^2$   \\
                 &                             & $D^{*0}\pi^0$     & $0.39903\,g_H^2$    & $D^{0}\pi^0$     & $0.31475\,g_H^2$   \\
                 &                             & $D^{*0}\eta$      & $0.07934\,g_H^2$    & $D^{0}\eta$      & $0.15758\,g_H^2$   \\  \hline

 $D_J^*(2760)$   & $1\,D\,\frac{3}{2}\,1^-$    & $D^{*+}\pi^-$     & $0.33541\,g_X^2$    & $D^{+}\pi^-$     & $1.28546\,g_X^2$    \\
                 &                             & $D_s^{*+}K^-$     & $0.06149\,g_X^2$    & $D_s^{+}K^-$     & $0.45974\,g_X^2$    \\
                 &                             & $D^{*0}\pi^0$     & $0.17110\,g_X^2$    & $D^{0}\pi^0$     & $0.65652\,g_X^2$    \\
                 &                             & $D^{*0}\eta$      & $0.08867\,g_X^2$    & $D^{0}\eta$      & $0.50190\,g_X^2$   \\ \hline

 $D_J(2740)$     & $1\,D\,\frac{3}{2}\,2^-$    &  $D^{*+}\pi^-$    & $0.87560\,g_X^2$    &                  &   \\
                 &                             &  $D_s^{*+}K^-$    & $0.13301\,g_X^2$    &                  &     \\
                 &                             &  $D^{*0}\pi^0$    & $0.44709\,g_X^2$    &                  &     \\
                 &                             &  $D^{*0}\eta$     & $0.20791\,g_X^2$    &                  &     \\ \hline

 $D_J(2740)$     & $1\,D\,\frac{5}{2}\,2^-$    & $D^{*+}\pi^-$     & $0.12739\,g_Y^2$    &                  &      \\
                 &                             & $D_s^{*+}K^-$     & $0.00193\,g_Y^2$    &                  &       \\
                 &                             & $D^{*0}\pi^0$     & $0.06597\,g_Y^2$    &                  &         \\
                 &                             & $D^{*0}\eta$      & $0.00524\,g_Y^2$    &                  &          \\ \hline

 $D_J^*(2760)$   & $1\,D\,\frac{5}{2}\,3^-$    & $D^{*+}\pi^-$     & $0.08910\,g_Y^2$    & $D^{+}\pi^-$     & $0.17382\,g_Y^2$        \\
                 &                             & $D_s^{*+}K^-$     & $0.00207\,g_Y^2$    & $D_s^{+}K^-$     & $0.01760\,g_Y^2$     \\
                 &                             & $D^{*0}\pi^0$     & $0.04607\,g_Y^2$    & $D^{0}\pi^0$     & $0.08995\,g_Y^2$      \\
                 &                             & $D^{*0}\eta$      & $0.00477\,g_Y^2$    & $D^{0}\eta$      & $0.02413\,g_Y^2$        \\ \hline    \hline
\end{tabular}
\end{center}
\caption{ The strong decay widths of the newly observed charmed
mesons with possible assignments.   }
\end{table}

\begin{table}
\begin{center}
\begin{tabular}{|c|c|cc|cc| }\hline\hline
                 & $n\,L\,s_\ell\,J^P$         & Decay channels    & Widths  [GeV]      &Decay channels     & Widths [GeV]   \\ \hline

 $D_J^*(3000)$   & $1\,F\,\frac{5}{2}\,2^+$    & $D^{*+}\pi^-$     & $0.35576\,g_Z^2$    & $D^{+}\pi^-$     & $1.03474\,g_Z^2$        \\
                 &                             & $D_s^{*+}K^-$     & $0.09260\,g_Z^2$    & $D_s^{+}K^-$     & $0.38738\,g_Z^2$       \\
                 &                             & $D^{*0}\pi^0$     & $0.18146\,g_Z^2$    & $D^{0}\pi^0$     & $0.52903\,g_Z^2$      \\
                 &                             & $D^{*0}\eta$      & $0.11279\,g_Z^2$    & $D^{0}\eta$      & $0.40754\,g_Z^2$       \\ \hline

 $D_J(3000)$     & $1\,F\,\frac{5}{2}\,3^+$    &  $D^{*+}\pi^-$    & $0.71619\,g_Z^2$    &                  &      \\
                 &                             &  $D_s^{*+}K^-$    & $0.16586\,g_Z^2$    &                  &      \\
                 &                             &  $D^{*0}\pi^0$    & $0.36571\,g_Z^2$    &                  &      \\
                 &                             &  $D^{*0}\eta$     & $0.21080\,g_Z^2$    &                  &      \\ \hline

 $D_J(3000)$     & $1\,F\,\frac{7}{2}\,3^+$    & $D^{*+}\pi^-$     & $1.70315\,g_R^2$    &                  &      \\
                 &                             & $D_s^{*+}K^-$     & $0.13668\,g_R^2$    &                  &      \\
                 &                             & $D^{*0}\pi^0$     & $0.87774\,g_R^2$    &                  &       \\
                 &                             & $D^{*0}\eta$      & $0.21081\,g_R^2$    &                  &      \\ \hline

 $D_J^*(3000)$   & $1\,F\,\frac{7}{2}\,4^+$    & $D^{*+}\pi^-$     & $1.25746\,g_R^2$    & $D^{+}\pi^-$     & $2.42276\,g_R^2$      \\
                 &                             & $D_s^{*+}K^-$     & $0.12371\,g_R^2$    & $D_s^{+}K^-$     & $0.43906\,g_R^2$       \\
                 &                             & $D^{*0}\pi^0$     & $0.64701\,g_R^2$    & $D^{0}\pi^0$     & $1.25012\,g_R^2$      \\
                 &                             & $D^{*0}\eta$      & $0.17952\,g_R^2$    & $D^{0}\eta$      & $0.52012\,g_R^2$       \\ \hline

 $D_J^*(3000)$   & $2\,P\,\frac{1}{2}\,0^+$    &                   &                     & $D^{+}\pi^-$     & $4.59878\,g_S^2$        \\
                 &                             &                   &                     & $D_s^{+}K^-$     & $3.76337\,g_S^2$       \\
                 &                             &                   &                     & $D^{0}\pi^0$     & $2.31591\,g_S^2$      \\
                 &                             &                   &                     & $D^{0}\eta$      & $2.99368\,g_S^2$       \\ \hline

 $D_J(3000)$     & $2\,P\,\frac{1}{2}\,1^+$    &  $D^{*+}\pi^-$    & $3.32572\,g_S^2$    &                  &      \\
                 &                             &  $D_s^{*+}K^-$    & $2.41575\,g_S^2$    &                  &      \\
                 &                             &  $D^{*0}\pi^0$    & $1.67433\,g_S^2$    &                  &      \\
                 &                             &  $D^{*0}\eta$     & $2.06738\,g_S^2$    &                  &      \\ \hline

 $D_J(3000)$     & $2\,P\,\frac{3}{2}\,1^+$    & $D^{*+}\pi^-$     & $2.73390\,g_T^2$    &                  &      \\
                 &                             & $D_s^{*+}K^-$     & $0.68819\,g_T^2$    &                  &      \\
                 &                             & $D^{*0}\pi^0$     & $1.38912\,g_T^2$    &                  &       \\
                 &                             & $D^{*0}\eta$      & $0.71469\,g_T^2$    &                  &      \\ \hline

 $D_J^*(3000)$   & $2\,P\,\frac{3}{2}\,2^+$    & $D^{*+}\pi^-$     & $1.91077\,g_T^2$    & $D^{+}\pi^-$     & $2.00996\,g_T^2$      \\
                 &                             & $D_s^{*+}K^-$     & $0.53862\,g_T^2$    & $D_s^{+}K^-$     & $0.79622\,g_T^2$       \\
                 &                             & $D^{*0}\pi^0$     & $0.97002\,g_T^2$    & $D^{0}\pi^0$     & $1.02155\,g_T^2$      \\
                 &                             & $D^{*0}\eta$      & $0.54072\,g_T^2$    & $D^{0}\eta$      & $0.71319\,g_T^2$       \\ \hline

$D_J(3000)$      & $3\,S\,\frac{1}{2}\,0^-$    & $D^{*+}\pi^-$     & $3.22680\,g_H^2$    &                   &   \\
                 &                             & $D^{*+}_sK^-$     & $1.43857\,g_H^2$    &                   &   \\
                 &                             & $D^{*0}\pi^0$     & $1.62798\,g_H^2$    &                   &   \\
                 &                             & $D^{*0}\eta$      & $1.22589\,g_H^2$    &                   &   \\  \hline

 $D^*_J(3000)$   & $3\,S\,\frac{1}{2}\,1^-$    & $D^{*+}\pi^-$     & $2.34605\,g_H^2$    & $D^{+}\pi^-$      & $1.49813\,g_H^2$        \\
                 &                             & $D^{*+}_sK^-$     & $1.11939\,g_H^2$    & $D_s^{+} K^-$     & $0.87742\,g_H^2$      \\
                 &                             & $D^{*0}\pi^0$     & $1.18299\,g_H^2$    & $D^{0}\pi^0$      & $0.75568\,g_H^2$      \\
                 &                             & $D^{*0}\eta$      & $0.93470\,g_H^2$    & $D^{0}\eta$       & $0.68341\,g_H^2$    \\  \hline

    \hline
\end{tabular}
\end{center}
\caption{ The strong decay widths of the newly observed charmed
mesons with possible assignments.   }
\end{table}

\begin{table}
\begin{center}
\begin{tabular}{|c|c|cc|cc| }\hline\hline
                 & $n\,L\,s_\ell\,J^P$         & Decay channels    & $\widehat{\Gamma}$  &Decay channels    & $\widehat{\Gamma}$   \\ \hline

 $D_J(2580)$     & $2\,S\,\frac{1}{2}\,0^-$    & $D^{*+}\pi^-$     & 1                   &                  &    \\
                 &                             & $D^{*+}_sK^-$     & 0                   &                  &   \\
                 &                             & $D^{*0}\pi^0$     & 0.51                &                  &   \\
                 &                             & $D^{*0}\eta$      & 0.02                &                  &   \\  \hline

 $D^*_J(2650)$   & $2\,S\,\frac{1}{2}\,1^-$    & $D^{*+}\pi^-$     & 1                   & $D^{+}\pi^-$     & 0.79   \\
                 &                             & $D^{*+}_sK^-$     & 0.04                & $D_s^{+} K^-$    & 0.20   \\
                 &                             & $D^{*0}\pi^0$     & 0.51                & $D^{0}\pi^0$     & 0.40   \\
                 &                             & $D^{*0}\eta$      & 0.10                & $D^{0}\eta$      & 0.20   \\  \hline

 $D_J^*(2760)$   & $1\,D\,\frac{3}{2}\,1^-$    & $D^{*+}\pi^-$     & 1                   & $D^{+}\pi^-$     & 3.83    \\
                 &                             & $D_s^{*+}K^-$     & 0.18                & $D_s^{+}K^-$     & 1.37    \\
                 &                             & $D^{*0}\pi^0$     & 0.51                & $D^{0}\pi^0$     & 1.96    \\
                 &                             & $D^{*0}\eta$      & 0.26                & $D^{0}\eta$      & 1.50   \\ \hline

 $D_J(2740)$     & $1\,D\,\frac{3}{2}\,2^-$    &  $D^{*+}\pi^-$    & 1                   &                  &   \\
                 &                             &  $D_s^{*+}K^-$    & 0.15                &                  &     \\
                 &                             &  $D^{*0}\pi^0$    & 0.51                &                  &     \\
                 &                             &  $D^{*0}\eta$     & 0.24                &                  &     \\ \hline

 $D_J(2740)$     & $1\,D\,\frac{5}{2}\,2^-$    & $D^{*+}\pi^-$     & 1                   &                  &      \\
                 &                             & $D_s^{*+}K^-$     & 0.02                &                  &       \\
                 &                             & $D^{*0}\pi^0$     & 0.52                &                  &         \\
                 &                             & $D^{*0}\eta$      & 0.04                &                  &          \\ \hline

 $D_J^*(2760)$   & $1\,D\,\frac{5}{2}\,3^-$    & $D^{*+}\pi^-$     & 1                   & $D^{+}\pi^-$     & 1.95        \\
                 &                             & $D_s^{*+}K^-$     & 0.02                & $D_s^{+}K^-$     & 0.20     \\
                 &                             & $D^{*0}\pi^0$     & 0.52                & $D^{0}\pi^0$     & 1.01      \\
                 &                             & $D^{*0}\eta$      & 0.05                & $D^{0}\eta$      & 0.27        \\ \hline    \hline
\end{tabular}
\end{center}
\caption{ The ratios $\widehat{\Gamma}=\frac{\Gamma}{\Gamma(D^{(*)}_J\to D^{*+}\pi^-)}$ of the strong decays of the newly observed charmed
mesons with possible assignments.   }
\end{table}

\begin{table}
\begin{center}
\begin{tabular}{|c|c|cc|cc| }\hline\hline
                 & $n\,L\,s_\ell\,J^P$         & Decay channels    & $\widehat{\Gamma}$  &Decay channels    & $\widehat{\Gamma}$   \\ \hline

 $D_J^*(3000)$   & $1\,F\,\frac{5}{2}\,2^+$    & $D^{*+}\pi^-$     & 1                   & $D^{+}\pi^-$     & 2.91       \\
                 &                             & $D_s^{*+}K^-$     & 0.26                & $D_s^{+}K^-$     & 1.09       \\
                 &                             & $D^{*0}\pi^0$     & 0.51                & $D^{0}\pi^0$     & 1.49     \\
                 &                             & $D^{*0}\eta$      & 0.32                & $D^{0}\eta$      & 1.15       \\ \hline

 $D_J(3000)$     & $1\,F\,\frac{5}{2}\,3^+$    &  $D^{*+}\pi^-$    & 1                   &                  &      \\
                 &                             &  $D_s^{*+}K^-$    & 0.23                &                  &      \\
                 &                             &  $D^{*0}\pi^0$    & 0.51                &                  &      \\
                 &                             &  $D^{*0}\eta$     & 0.29                &                  &      \\ \hline

 $D_J(3000)$     & $1\,F\,\frac{7}{2}\,3^+$    & $D^{*+}\pi^-$     & 1                   &                  &      \\
                 &                             & $D_s^{*+}K^-$     & 0.08                &                  &      \\
                 &                             & $D^{*0}\pi^0$     & 0.52                &                  &       \\
                 &                             & $D^{*0}\eta$      & 0.12                &                  &      \\ \hline

 $D_J^*(3000)$   & $1\,F\,\frac{7}{2}\,4^+$    & $D^{*+}\pi^-$     & 1                   & $D^{+}\pi^-$     & 1.93      \\
                 &                             & $D_s^{*+}K^-$     & 0.10                & $D_s^{+}K^-$     & 0.35       \\
                 &                             & $D^{*0}\pi^0$     & 0.51                & $D^{0}\pi^0$     & 0.99      \\
                 &                             & $D^{*0}\eta$      & 0.14                & $D^{0}\eta$      & 0.41       \\ \hline

 $D_J^*(3000)$   & $2\,P\,\frac{1}{2}\,0^+$    &                   &                     & $D^{+}\pi^-$     & 1       \\
                 &                             &                   &                     & $D_s^{+}K^-$     & 0.82       \\
                 &                             &                   &                     & $D^{0}\pi^0$     & 0.50      \\
                 &                             &                   &                     & $D^{0}\eta$      & 0.65       \\ \hline

 $D_J(3000)$     & $2\,P\,\frac{1}{2}\,1^+$    &  $D^{*+}\pi^-$    & 1                   &                  &      \\
                 &                             &  $D_s^{*+}K^-$    & 0.73                &                  &      \\
                 &                             &  $D^{*0}\pi^0$    & 0.50                &                  &      \\
                 &                             &  $D^{*0}\eta$     & 0.62                &                  &      \\ \hline

 $D_J(3000)$     & $2\,P\,\frac{3}{2}\,1^+$    & $D^{*+}\pi^-$     & 1                   &                  &      \\
                 &                             & $D_s^{*+}K^-$     & 0.25                &                  &      \\
                 &                             & $D^{*0}\pi^0$     & 0.51                &                  &       \\
                 &                             & $D^{*0}\eta$      & 0.26                &                  &      \\ \hline

 $D_J^*(3000)$   & $2\,P\,\frac{3}{2}\,2^+$    & $D^{*+}\pi^-$     & 1                   & $D^{+}\pi^-$     & 1.05      \\
                 &                             & $D_s^{*+}K^-$     & 0.28                & $D_s^{+}K^-$     & 0.42       \\
                 &                             & $D^{*0}\pi^0$     & 0.51                & $D^{0}\pi^0$     & 0.53      \\
                 &                             & $D^{*0}\eta$      & 0.28                & $D^{0}\eta$      & 0.37       \\ \hline

$D_J(3000)$      & $3\,S\,\frac{1}{2}\,0^-$    & $D^{*+}\pi^-$     & 1                   &                   &   \\
                 &                             & $D^{*+}_sK^-$     & 0.45                &                   &   \\
                 &                             & $D^{*0}\pi^0$     & 0.50                &                   &   \\
                 &                             & $D^{*0}\eta$      & 0.38                &                   &   \\  \hline

 $D^*_J(3000)$   & $3\,S\,\frac{1}{2}\,1^-$    & $D^{*+}\pi^-$     & 1                   & $D^{+}\pi^-$      & 0.64        \\
                 &                             & $D^{*+}_sK^-$     & 0.48                & $D_s^{+} K^-$     & 0.37      \\
                 &                             & $D^{*0}\pi^0$     & 0.50                & $D^{0}\pi^0$      & 0.32      \\
                 &                             & $D^{*0}\eta$      & 0.40                & $D^{0}\eta$       & 0.29   \\  \hline

    \hline
\end{tabular}
\end{center}
\caption{ The ratios $\widehat{\Gamma}=\frac{\Gamma}{\Gamma(D^{(*)}_J\to D^{*+}\pi^-)}$ of the strong decays of the newly observed charmed
mesons with possible assignments.  In the decays $0^+_{\frac{1}{2}} \to 0^-_{\frac{1}{2}}+\mathcal{P}$, $\widehat{\Gamma}=\frac{\Gamma}{\Gamma(D^{*}_J\to D^{+}\pi^-)}$. }
\end{table}

In Table 8, we present the experimental data on the ratio $\frac{\Gamma(D_2^*(2460)\to D^{+}\pi^-)}{\Gamma(D_2^*(2460)\to D^{*+}\pi^-)}$ for the well
established meson $D_2^*(2460)$ from the BaBar \cite{Babar2010}, CLEO \cite{CLEO1994,CLEO1990}, ARGUS \cite{ARGUS1989}, and ZEUS
\cite{ZEUS2009} collaborations,  the present prediction $2.29$ based on the heavy meson effective theory in the leading order approximation is
in excellent agreement with the average experimental value $2.35$. The heavy meson  effective theory in the leading
order approximation works well.

\begin{table}
\begin{center}
\begin{tabular}{|c|c|c|c|c|c|c| }\hline\hline
                 BaBar    &  CLEO               & CLEO           & ARGUS&ZEUS          &This work\\ \hline
  $1.47\pm0.03\pm0.16$    & $2.2\pm0.7\pm0.6$   & $2.3\pm 0.8$   & $3.0\pm1.1\pm1.5$   &$2.8\pm 0.8^{+0.5}_{-0.6}$ &$2.29$\\ \hline
       \hline
\end{tabular}
\end{center}
\caption{ The  experimental values of the ratio $\frac{\Gamma\left(D^*_2(2460)^0\to
D^+\pi^-\right)}{\Gamma\left(D^*_2(2460)^0\to D^{*+}\pi^-\right)} $
  compared to the prediction of the leading order heavy meson effective theory. }
\end{table}

If we saturate the total widths of the doublet $(D_J(2580),D_J^*(2650))$ or $(D(2550),D(2600))$ with the two-body decays to the ground states $(D,D^*)_{\frac{1}{2}}+\mathcal{P}$,
the total  widths   $\Gamma_{D_J(2580)}\approx 1.3g_H^2\,\rm{GeV}$, $\Gamma_{D^*_J(2650)}\approx 2.5g_H^2\,\rm{GeV}$, $\Gamma_{D(2550)}\approx 1.7g_H^2\,\rm{GeV}$, $\Gamma_{D(2600)}\approx 2.0g_H^2\,\rm{GeV}$, the ratio
$\frac{\Gamma_{D_J(2580)}}{\Gamma_{D^*_J(2650)}}\approx0.52$  is smaller than the experimental data
$\frac{\Gamma_{D_J(2580)}}{\Gamma_{D^*_J(2650)}}\approx 1.27$ from the LHCb collaboration \cite{LHCb1307},
 the ratio
$\frac{\Gamma_{D(2550)}}{\Gamma_{D(2600)}}\approx0.85$  is also smaller than the experimental data $\frac{\Gamma_{D(2550)}}{\Gamma_{D(2600)}}\approx1.40$   from the BaBar collaboration \cite{Babar2010}.
The $2\,\rm{S}$ states $(D_J(2580),D_J^*(2650))$ or $(D(2550),D(2600))$ can decay to the $\rm{P}$-wave states $(0^+,1^+)_{\frac{1}{2}}+\mathcal{P}$ and $(1^+,2^+)_{\frac{3}{2}}+\mathcal{P}$, which also contribute to the total decay widths. While the decays to  the $\rm{P}$-wave states
$(0^+,1^+)_{\frac{1}{2}}+V$ are kinematically forbidden,   where the $V$ denotes the light vector mesons $\rho$, $K^*$, $\omega$.
If those contributions are taken into account properly, the discrepancy may be smeared. The  ratio $\frac{\Gamma_{D_1(2420)}}{\Gamma_{D_2^*(2460)}}\approx0.31$ from the heavy meson effective theory in the leading order
approximation is also smaller than  the experimental data $0.81$ \cite{LHCb1307}.
For the charmed mesons, the flavor and spin violation corrections of order $\mathcal
{O}(1/m_Q)$  may be sizable, we have to introduce new unknown coupling constants,  the discrepancy may be
smeared with the optimal parameters \cite{Falk1996}, furthermore,  more precise
measurements are needed to make a reliable comparison.

\section{Conclusion}
In this article, we tentatively identify the  charmed mesons $D_J(2580)$, $D_J^*(2650)$, $D_J(2740)$,
$D^*_J(2760)$,   $D_J(3000)$, $D_J^*(3000)$  observed by the LHCb collaboration according to their spin, parity and masses, then study their strong decays
 to the ground state charmed mesons plus light pseudoscalar mesons with
the heavy meson  effective theory in the leading order approximation, and obtain explicit expressions of the decay  widths.
The strong coupling constants $g_H$, $g_S$, $g_T$, $g_X$, $g_Y$, $g_Z$, $g_R$ in the heavy meson chiral Lagrangians  can be fitted to the experimental
data in the future at the LHCb,  BESIII, KEK-B, $\rm{\bar{P}ANDA}$, and be taken as basic input parameters in studying the interactions among the heavy mesons.
While the ratios among the decay widths can be used to confirm or reject the assignments of the newly observed charmed mesons.

\section*{Acknowledgment}
This  work is supported by National Natural Science Foundation of
China, Grant Numbers 11075053, 11375063, and the Fundamental Research Funds for the Central Universities.

\end{document}